\RequirePackage{arydshln}
\documentclass[aps,twocolumn,nofootinbib,superscriptaddress,preprintnumbers,pra,10pt,floatfix]{revtex4-1}

\usepackage[utf8]{inputenc}

\usepackage{float}
\usepackage{amsmath,amssymb}
\usepackage{dsfont} 
\usepackage{hyperref}
\usepackage{graphicx}
\usepackage{enumitem}
\usepackage{mathtools}
\usepackage{bbold}
\usepackage{multirow}
\usepackage{ytableau}
\usepackage{youngtab}
\usepackage{braket}
\usepackage{soul}
\usepackage{cancel}
\usepackage{xcolor}
\usepackage[normalem]{ulem}
\usepackage[export]{adjustbox}

\usepackage{fontawesome}

\usepackage{bbm}
\usepackage{mathrsfs}

\usepackage{lipsum}

\usepackage{colortbl} 
\definecolor{Gray}{gray}{0.95}
\definecolor{RGray}{gray}{0.290}
\definecolor{CGray}{gray}{0.92}

\usepackage{arydshln}

\definecolor{blueLinks}{RGB}{0,84,159}
\newcommand{\myColor}[0]{blueLinks}
\hypersetup{
	colorlinks, 
	bookmarksopen, 
	bookmarksnumbered,
	pdftoolbar=false, 
	pdfmenubar=false, 
	pdffitwindow=false, 
	pdfstartview={FitH},	 
	citecolor=\myColor, 
	linkcolor=\myColor,	
	urlcolor=\myColor, 
	pdftitle={On the EFT validity for Drell-Yan tails at the LHC}, 
	pdfsubject={EFT validity - Drell-Yan}, 
	pdfnewwindow=true, 
	linktoc=all
}

\makeatletter
\g@addto@macro\bfseries{\boldmath}
\makeatother

\makeatletter
\renewcommand\paragraph{\@startsection{paragraph}{4}{\z@}%
                                    {3.25ex \@plus1ex \@minus.2ex}%
                                    {-1em}%
                                    {\normalfont\normalsize\bfseries}}
\makeatother

\allowdisplaybreaks
\begin{document}

\preprint{TTK-24-53}
\preprint{P3H-24-094}
\preprint{DESY-24-217}

\title{ On the EFT validity for Drell-Yan tails at the LHC}

\author{L.~Allwicher}
\email{lukas.allwicher@desy.de}
\affiliation{Deutsches Elektronen-Synchrotron DESY, Notkestr.~85, 22607 Hamburg, Germany}
\author{D.~A.~Faroughy}
\email{darius.faroughy@rutgers.edu}
\affiliation{NHETC, Department of Physics and Astronomy, Rutgers University, Piscataway, NJ 08854, USA}
\author{M.~Martines}
\email{matheus.martines.silva@usp.br}
\affiliation{Instituto de Física, Universidade de São Paulo, São Paulo – São Paulo 05580-090, Brazil.}
\affiliation{IJCLab, P\^ole Th\'eorie (Bat.~210), CNRS/IN2P3 et Universit\'e Paris-Saclay, 91405 Orsay, France}
\author{O.~Sumensari}
\email{olcyr.sumensari@ijclab.in2p3.fr}
\affiliation{IJCLab, P\^ole Th\'eorie (Bat.~210), CNRS/IN2P3 et Universit\'e Paris-Saclay, 91405 Orsay, France}
\author{F.~Wilsch}
\email{felix.wilsch@physik.rwth-aachen.de}
\affiliation{Institute for Theoretical Particle Physics and Cosmology, RWTH Aachen University, \\ Sommerfeldstr.~16, D-52074 Aachen, Germany}

\begin{abstract}
\vspace{5mm}
In this article, we examine the validity range of the Effective Field Theory (EFT) description of high-energy Drell-Yan processes at the LHC. 
To this purpose, we consider explicit mediators that contribute to these processes in the $s$- and $t$-channels, comparing their effects in Drell-Yan distributions with the ones obtained by matching onto the corresponding EFT. 
We determine the conditions for the EFT results to accurately describe these scenarios.
In particular, we explore the impact of including dimension-eight~$(d=8)$ operators in the faster convergence of the EFT series, at the analytical and numerical level, considering contributions to the cross section up to the square of $d=8$ EFT operator insertions.
Moreover, we discuss the possible implications of clipping LHC data and illustrate results for a specific New-Physics scenario motivated by low-energy flavor data.
\vspace{3mm}
\end{abstract}

\maketitle

\allowdisplaybreaks

\newcommand{\cmd}[1]{{\texttt{#1}}\xspace}
\newcommand{\dd}{\mathop{}\!\mathrm{d}}
\newcommand{\braces}[1]{\left\lbrace #1 \right\rbrace}
\newcommand{\brackets}[1]{\left( #1 \right)}
\newcommand{\squarebrackets}[1]{\left[ #1 \right]} 
\newcommand{\angles}[1]{\left\langle #1\right\rangle}
\newcommand{\abs}[1]{\left\lvert #1 \right\rvert}
\newcommand{\norm}[1]{\left\Vert #1 \right\Vert}
\renewcommand{\Im}{\mathop{\mathrm{Im}}}
\renewcommand{\Re}{\mathop{\mathrm{Re}}}
\renewcommand{\L}{\mathcal{L}}
\newcommand{\ord}[1]{\mathcal{O}\!\left( #1 \right)}
\newcommand{\ineqgraphics}[1]{\vcenter{\hbox{\includegraphics[]{#1}}}} 

\newcommand{\SMEFT}[0]{\text{SM\,EFT}\xspace}
\newcommand{\vev}[0]{vacuum expectation value\xspace}
\newcommand{\QFT}[0]{quantum field theory\xspace}
\newcommand{\SU}[1]{\mathrm{SU}(#1)}
\newcommand{\C}[2]{C_{\underset{#2}{#1}}}
\newcommand{\vT}[0]{\overline{v}_T}

\newcommand{\cO}{{\mathcal O}}
\newcommand{\cQ}{{\mathcal Q}}
\newcommand{\cC}{{\mathcal C}}
\newcommand{\cL}{{\mathcal L}}
\newcommand{\cH}{{\mathcal H}}
\newcommand{\cI}{{\mathcal I}}
\newcommand{\cA}{{\mathcal A}}
\newcommand{\cB}{{\mathcal B}}
\newcommand{\cF}{{\mathcal F}}
\newcommand{\cP}{{\mathcal P}}
\newcommand{\cS}{{\mathcal S}}
\newcommand{\cT}{{\mathcal T}}
\newcommand{\cU}{{\mathcal U}}
\newcommand{\cN}{{\mathcal N}}
\newcommand{\cM}{{\mathcal M}}
\newcommand{\cZ}{{\mathcal Z}}

\newcommand{\SMrep}[3]{({\bf #1},\,{\bf #2},\,#3)}
\newcommand{\SMrepbar}[3]{({\bf \bar#1},\,{\bf #2},\,#3)}

\definecolor{myGreen}{rgb}{0.0, 0.7, 0.1}

\newcommand{\dario}[1]{{\color{blue} [OS:#1]}}
\newcommand{\olcyr}[1]{{\color{orange} [OS:#1]}}
\newcommand{\felix}[1]{{\color{myGreen} [FW: #1]}}
\newcommand{\felixx}[1]{{\color{myGreen} #1}}
\newcommand{\lukas}[1]{{\color{purple}[L.A.: #1]}}
\newcommand{\lukkas}[1]{{\color{purple} #1}}
\newcommand{\matheus}[1]{{\color{blue} [MM:#1]}}


\section{Introduction}

The absence of signals in direct searches for new particles at the Large Hadron Collider~(LHC) indicates that there may be a separation between the electroweak scale and the unknown scale of New Physics. With this assumption, the most convenient approach to describe LHC data becomes Effective Field Theories~(EFTs), and the main target of LHC searches are deviations from the Standard Model~(SM) expectations in the high-energy tails of the kinematical distributions. Such non-resonant analyses have been performed for many channels, including the Drell-Yan processes $pp\to\ell\ell$ and $pp\to\ell\nu$ ($\ell=e,\mu,\tau$), which are efficient probes of semileptonic four-fermion operators.

The importance of the high-$p_T$ tails of the Drell-Yan distributions can be understood considering the naive EFT predictions from contact interactions to high-energy amplitudes, which can scale as $(E/\Lambda)^n$ for $E \ll \Lambda$, with $n>0$, where $\Lambda$ denotes the EFT cutoff and $E$ is the typical energy scale of these processes. Therefore, if the EFT description is valid, the energy enhancement of the Drell-Yan cross section allows us to derive stringent constraints on the corresponding Wilson coefficients, which can be competitive, e.g., to electroweak observables~\cite{Farina:2016rws,deBlas:2013qqa,Dawson:2018dxp,Torre:2020aiz}, as well as to low-energy flavor bounds, if we exploit the flavor content of the proton~\cite{Allwicher:2022gkm,Allwicher:2022mcg,Cirigliano:2012ab,Chang:2014iba,Faroughy:2016osc,Greljo:2018tzh,Cirigliano:2018dyk,Marzocca:2020ueu,Iguro:2020keo,Fuentes-Martin:2020lea,Greljo:2017vvb,Angelescu:2020uug,Endo:2021lhi,Jaffredo:2021ymt,Crivellin:2021rbf,Greljo:2022jac,Bressler:2022zhv,Boughezal:2023nhe,Descotes-Genon:2023pen,Grunwald:2023nli,Bartocci:2023nvp,Hiller:2024vtr,Becirevic:2024pni}. 

The main caveat of EFT analyses at colliders is that the experimental sensitivity on $\mathcal{C}/\Lambda^2$ is not always sufficient to consistently probe EFTs, i.e.,~with $E \ll \Lambda$, while having Wilson coefficients~$\mathcal{C}$ within the perturbative regime~\cite{Brivio:2022pyi}. 
The applicability range of an EFT and the impact of $d=8$ operators can be estimated in various manners~\cite{Contino:2016jqw,Azatov:2016sqh,Cohen:2021gdw,Brivio:2022pyi,Corbett:2024evt}. 
For instance,~by defining the so-called maximal cutoff scale~($\Lambda_\mathrm{max}$) that can be consistently probed for a perturbative scenario. 
The value of~$\Lambda_\mathrm{max}$ is defined by the requirement that the $2\to 2$ scattering amplitude does not exceed the $16\pi^2$~limit arising from perturbativity~\cite{Farina:2016rws}, giving a first indication of the limits of the EFT description. 
Another approach often used is to provide constraints as a function of a mass scale~$M_{\text{cut}}$ in some kinematical variable relevant to the process considered, above which all data are discarded~\cite{Contino:2016jqw,Farina:2016rws}. 
In this way, it is possible to extract weaker, but potentially more robust, EFT limits, even though the phase-space migration of events in the detector simulation can be a potential problem~\cite{Falkowski:2016cxu,Brivio:2022pyi}.

The approaches described above are helpful for naively assessing the limits of applicability of the constraints derived by the EFT. However, the definite answer to whether EFT results can be applied to a given scenario will depend on its ultraviolet~(UV) properties, in particular, if there is a small separation between~$E$ and~$\Lambda$, as it is often the case in collider studies. For instance, tree-level contributions in the $s$- or $t$-channels will deform the EFT bounds differently when $E$ approaches the EFT cutoff. While non-resonant $t$-channel contributions to Drell-Yan processes by leptoquark states~\cite{Buchmuller:1986zs,Dorsner:2016wpm} can reproduce the EFT results even for values of~$E$ in the vicinity of~$\Lambda$, it is well-known that the EFT description fails for resonant colorless $s$-channel mediators already for $E$ much below~$\Lambda$~\cite{Farina:2016rws}.~\footnote{The situation is even less intuitive for loop-level contributions, as shown, e.g.,~in Ref.~\cite{Paradisi:2022vqp} for probes of effective dipole operators in a high-energy lepton collider.} 
The comparison of the direct limits on these concrete scenarios with the ones derived through the EFT approach allows us to precisely determine which mediator masses are well-described by the EFT. This will depend on several factors: (i)~the topology of the diagram, (ii)~the flavor of the initial quarks and the corresponding Parton-Distribution-Function~(PDF) suppression, and (iii)~the sensitivity of the experimental search, which depends on the flavor of all the particles considered.

The purpose of this paper is to assess the range of validity of Drell-Yan constraints on EFTs by \emph{directly comparing} limits on selected concrete UV~scenarios with the results derived through EFT analyses. We will consider two tree-level benchmark scenarios, classified in terms of their $(SU(3)_c, SU(2)_L,U(1)_Y)$ quantum numbers: (i)~a vector leptoquark $U_1\sim (\mathbf{3},\mathbf{1},2/3)$ and (ii)~a colorless $Z^\prime\sim (\mathbf{1},\mathbf{1},0)$ boson; which contribute to these processes in the $t$- and $s$-channel, respectively. We will consider the neutral-current dilepton searches made available by CMS~\cite{CMS:2021ctt} and ATLAS~\cite{ATLAS:2020zms}, which have been recast in Ref.~\cite{Allwicher:2022gkm,Allwicher:2022mcg}, and we will perform this comparison for different flavors of the initial and final states. Our main goal is to establish a clear prescription for the situations where the EFT results can be applied to concrete UV~scenarios by simply matching the models to the EFT Lagrangian, thus avoiding costly and model-dependent numerical simulations of the mediator's propagation. The impact of dimension-eight operators in our analysis will also be explored at the amplitude and cross-section level (see also Ref.~\cite{Allwicher:2022gkm,Boughezal:2022nof,Boughezal:2021tih,Dawson:2023ebe}).

The remainder of this article is organized as follows. In Sec.~\ref{sec:drell-yan}, we introduce our framework and describe the Drell-Yan probes of flavor at high-$p_T$. In Sec.~\ref{sec:concrete}, we consider a few concrete UV~scenarios that are matched onto the SMEFT at tree level. We then study in Sec.~\ref{sec:eft-vs-models} the convergence of the EFT expansion and in Sec.~\ref{sec:clipping} the impact of clipping data. In Sec.~\ref{sec:numerical}, we illustrate our results for a specific realization of the $U_1$~leptoquark model, which has been proposed to accommodate anomalies in $B$-physics data. We summarize our main findings in Sec.~\ref{sec:summary}. 

\section{EFT APPROACH}
\label{sec:drell-yan}

We start by defining our framework.  We consider the SMEFT Lagrangian~\cite{Buchmuller:1985jz,Grzadkowski:2010es}, which is invariant under the $SU(3)_c \times SU(2)_L \times U(1)_Y$ gauge symmetry, and we keep operators up to dimension $d=8$,
\begin{align}
\label{eq:smeft}
    \mathcal{L}_{\mathrm{SMEFT}} \supset \sum_a \dfrac{\mathcal{C}_a^{(6)}}{\Lambda^2} \mathcal{O}_a^{(6)} + \sum_a \dfrac{\mathcal{C}^{(8)}_a}{\Lambda^4} \mathcal{O}^{(8)}_a+\dots\,,
\end{align}
where $\Lambda$ denotes the EFT cutoff. The effective coefficients are generically denoted by~$\smash{\mathcal{C}_a^{(d)}}$ and the effective operators~$\smash{\mathcal{O}_a^{(d)}}$ can be of several types and with different flavor content, which are labeled by the index~$a$. We will consider the Warsaw basis for the $d=6$ operators~\cite{Grzadkowski:2010es} and its extension to $d=8$ from Ref.~\cite{Murphy:2020rsh} (see also Ref.~\cite{Li:2020gnx}), which are both implemented in the {\tt HighPT} package~\cite{Allwicher:2022mcg}. For convenience, the operators appearing in the benchmark scenarios that will be discussed in the following are collected in Table~\ref{tab:SMEFT-ope}.

\paragraph*{Amplitude decomposition} The most general decomposition of the four-point scattering amplitude for the $\bar{q}_iq_j\to \ell_\alpha^- \ell_\beta^+$ process (with $q=u,d$), which is Lorentz invariant and consistent with the $SU(3)_c\times U(1)_{\rm em}$ gauge symmetry, reads  
\begin{equation}
\label{eq:amplitude}
\mathcal{A}(\bar{q}_i q_j \to {\ell}_{\alpha}^-\ell^+_{\beta})= \dfrac{1}{v^2} \sum_{\Gamma \otimes \Gamma^\prime} \mathcal{F}_{\Gamma,\Gamma^\prime}^{\alpha \beta i j} \, \big{(} \bar{\ell}_\alpha \Gamma \ell_\beta \big{)} \big{(} \bar{q}_i \Gamma^\prime q_j \big{)} \,,
\end{equation}
which is weighted by the Higgs vacuum expectation value $v=(\sqrt{2} G_F)^{-1/2}$, and quark and lepton flavor indices are denoted by Latin letters ($i,j=1,2,3$) and Greek letters ($\alpha,\beta=1,2,3$), respectively. The form factors $\smash{\mathcal{F}_{\Gamma,\Gamma^\prime}^{\alpha \beta i j} \equiv \mathcal{F}_{\Gamma,\Gamma^\prime}^{\alpha \beta i j}  (\hat{s},\hat{t})}$ are functions of the partonic Mandelstam variables $\hat{s}=(p_{q_j}+p_{\bar{q}_i})^2$ and $\hat{t}=(p_{q_j}-p_{\ell_\alpha})^2$, which describe the effects of EFT operators and/or concrete mediators, in addition to the SM contributions~\cite{Allwicher:2022gkm}. The viable Lorentz structures are given by 
\begin{equation}
\label{eq:FF-Lorentz}
\Gamma \otimes \Gamma^\prime \in \lbrace P_X \otimes P_Y\,, \gamma_\mu P_X \otimes \gamma^\mu P_Y \,, \sigma_{\mu\nu} P_X \otimes \sigma^{\mu\nu} P_X \rbrace\,,
\end{equation}
where $P_{X/Y}$ are the chirality projectors with $X,Y\in \lbrace L,R \rbrace$, in addition to dipoles that induce a milder energy-enhancement, cf.~Ref.~\cite{Allwicher:2022gkm}. In the SM, the only non-vanishing form factors are
\begin{equation}
\mathcal{F}_{V_X, V_Y}^{\alpha\beta ij} \equiv \delta_{\alpha\beta}\delta_{ij}\, \mathcal{F}_{XY}^{\mathrm{SM}}
\end{equation}
where we use the subscript $V_X$ for the vector current ($\gamma_\mu P_X$), and~\footnote{Similarly, we will denote the scalar ($P_X$) and tensor ($\sigma_{\mu\nu} P_X$) currents by $S_X$ and $T_{X}$, respectively.} 
\begin{equation}
\label{eq:F-SM}
\mathcal{F}_{XY}^{\mathrm{SM}} = \dfrac{v^2}{\hat{s}} e^2 Q_\ell Q_q + \dfrac{v^2}{\hat{s}-m_{Z}^2+i m_Z \Gamma_Z} g_\ell^X g_q^Y\,, 
\end{equation}
where $m_Z$ and $\Gamma_Z$ are the $Z$-boson mass and width, $g_\psi^X = (g_2/c_W) \,(t_{\psi_X}^{3}-s_W^2\,Q_\psi)$ denotes the $Z$~couplings to fermions $\psi \in \lbrace u,d,\ell \rbrace$, $Q_\psi$~stands for their electric charge, $t_{\psi_X}^{3}$~is the third component of the weak isospin, and $c_{W}\equiv \cos \theta_W$ and $s_{W}\equiv \sin \theta_W$, where $\theta_W$~denotes the weak mixing angle. In the high-energy limit $\hat{s} \gg m_Z$, the SM form factor behaves as $\mathcal{F}_{XY}^{\mathrm{SM}} \propto v^2/\hat{s}$. The matching between the form factors defined in Eq.~\eqref{eq:amplitude} and the $d\leq 8$ operators in the SMEFT as well as concrete mediators is given in Ref.~\cite{Allwicher:2022gkm}. Notice, in particular, that the new mediators will induce poles in $\hat{s}$, $\hat{t}$ or $\hat{u}=-\hat{s}-\hat{t}$, depending on the topology of the diagrams.

\paragraph*{Drell-Yan processes} The general amplitude defined in Eq.~\eqref{eq:amplitude} can be used to compute the partonic cross section for $\bar{q}_i q_j\to \ell_\alpha^-\ell_\beta^+$~\cite{Allwicher:2022gkm}, 
\begin{equation}
\label{eq:xsection-qq-ll}
\hspace{-0.35em}\hat{\sigma}(\bar{q}_i q_j \to \ell_\alpha^- \ell_\beta^+) = \dfrac{1}{48\pi v^4} \sum_{\Gamma \otimes \Gamma^\prime } \int_{-\hat{s}}^0 \mathrm{d}\hat{t}\, \mathcal{M}_{\Gamma, \Gamma^\prime} \, \Big{|}\mathcal{F}_{\Gamma,\Gamma^\prime}^{\alpha \beta i j}\Big{|}^2\,,
\end{equation}
where $\Gamma \otimes \Gamma^\prime$ can be the allowed Lorentz structures in Eq.~\eqref{eq:FF-Lorentz} and the weights $\mathcal{M}_{\Gamma, \Gamma^\prime} \equiv \mathcal{M}_{\Gamma, \Gamma^\prime}(\omega)$ are functions of $\omega = \hat{t}/\hat{s}$,
\begin{align}
\mathcal{M}_{V_X, V_Y} (\omega) &\equiv (1+2\omega)\,\delta_{XY}+\omega^2\,, \nonumber \\[0.3em] 
\mathcal{M}_{S_X, S_Y} (\omega) &\equiv 1/4\,, \\[0.3em] 
\mathcal{M}_{T_X, T_X} (\omega) &\equiv 4(1+2\omega)^2\,, \nonumber
\end{align}

\noindent where fermion masses are neglected.~\footnote{Notice that there are, in principle, off-diagonal terms between scalar and tensor form factors, but these vanish upon the integration over $\hat{t}\in (-\hat{s},0)$, if the form-factor dependence on~$\hat{t}$ is neglected. See Ref.~\cite{Allwicher:2022gkm} for the full expressions.} The contributions of $d=6$ operators in the SMEFT are such that the corresponding form factors~$\mathcal{F}^{(d=6)}$ are independent of~$\hat{t}$. Therefore, the integral over $\hat{t}\in(-\hat{s},0)$ can be explicitly performed in this case,
\begin{equation}
\hat{\sigma}(\bar{q}_i q_j \to \ell_\alpha^- \ell_\beta^+)_{\mathrm{EFT}} \overset{(d=6)}{=} \dfrac{\hat{s}}{48\pi v^4} \sum_{\Gamma \otimes \Gamma^\prime } {\overline{\mathcal{M}}}_{\Gamma, \Gamma^\prime} \Big{|}\mathcal{F}_{\Gamma,\Gamma^\prime}^{\alpha \beta i j}\Big{|}^2\,,
\end{equation}
where the integrated weight-factors now read
\begin{align}
\overline{\mathcal{M}}_{V_X,V_Y} &= 1/3 \,,\qquad
\overline{\mathcal{M}}_{S_X,S_Y} = 1/4\,,\nonumber\\[0.3em]
\overline{\mathcal{M}}_{T_X,T_X} &= 4/3\,.
\end{align}
The energy enhancement of the cross section is now explicit in the overall $\hat{s}/v^4$~pre-factors, since the~$\mathcal{F}^{(d=6)}$ form-factors converge to a constant value at large $\hat{s}$ for contact interactions~\cite{Allwicher:2022gkm}.

The LHC cross section can be generally written as a sum over all possible combinations of incoming quark flavors,
\begin{equation}
\label{eq:xsection-pp-ll}
\sigma(pp\to \ell_\alpha^-\ell_\beta^+) =  \sum_{i,j} \int \dfrac{\mathrm{d}\hat{s}}{s} \mathcal{L}_{\bar{q}_i q_j}\, \hat{\sigma}(\bar{q}_i q_j \to \ell_\alpha^- \ell_\beta^+) \,,
\end{equation}
with
\begin{equation}
\mathcal{L}_{ \bar{q_i} q_j}(\hat{s}) \equiv \int_{\hat{s}/s}^1 \dfrac{\mathrm{d}x}{x}\Big{[} f_{\bar{q}_i} (x,\mu_F) f_{{q}_j} \Big(\frac{\hat{s}}{s x},\mu_F\Big)+(\bar{q}_i \leftrightarrow q_j)\Big{]}\,,
\end{equation}
where $\sqrt{s}=13~\mathrm{TeV}$, $f_{{q}_j}$~and $f_{\bar{q}_i}$ denote the PDFs of $q_j$ and $\bar{q}_i$~quarks, and $\mu_F$~stands for the factorization scale. In our calculations, we set $\mu_F=\sqrt{\hat{s}}$ to the scale of the hard scattering process.

\paragraph*{LHC limits} We consider the neutral-current Drell-Yan constraints on the SMEFT provided by the {\tt HighPT} package~\href{https://highpt.github.io}{\faicon{github}}~\cite{Allwicher:2022gkm,Allwicher:2022mcg}. These results have been obtained through an appropriate recast of the relevant CMS~\cite{CMS:2021ctt} and ATLAS~\cite{ATLAS:2020zms} searches. More specifically, the event samples were generated with {\tt MadGraph5}~\cite{Alwall:2014hca}, showered and hadronized by {\tt Pythia8}~\cite{Sjostrand:2014zea}, and the final-state object reconstruction and detector simulations were performed using {\tt Delphes3}~\cite{deFavereau:2013fsa}, tuned to match the experimental searches. The {\tt PDF4LHC15\_nnlo\_mc} PDF set~\cite{Butterworth:2015oua} has been used in these reinterpretations. The final results are combined in a $\chi^2$-distribution, with the background estimates taken from the experimental papers (which include higher-order QCD corrections) and the New-Physics contributions calculated at tree level following the pipeline described above, see~Ref.~\cite{Allwicher:2022gkm,Allwicher:2022mcg} for more details.

\section{Benchmark scenarios }
\label{sec:concrete}

In this Section, we introduce the two benchmark scenarios that will be used for the comparison between the EFT and concrete models. These scenarios will be classified in terms of their SM quantum numbers $(SU(3)_c,SU(2)_L,U(1)_Y)$:

\paragraph*{$Z'\sim ({\bf 1}, {\bf 1}, 0)$\,:}
The first scenario that we consider is a gauge-singlet vector field that couples to the SM particles via the following Lagrangian
\begin{align}
    \label{eq:Zprime}
    \cL_{Z^\prime}= -\dfrac{1}{4} Z^\prime_{\mu\nu} Z^{\prime\,\mu\nu} + \dfrac{m_{Z^\prime}^2}{2} Z_\mu^{\prime} Z^{\prime\,\mu} + J^\mu Z_\mu^\prime \,, 
\end{align}
where ${Z^\prime}_{\mu\nu}=D_\mu Z^\prime_\nu - D_\nu Z^\prime_\mu$. For the $Z^\prime$~model considered here, we have $D_\mu Z^\prime = \partial_\mu Z^\prime$.~\footnote{The covariant derivative of the SM gauge group acting on a generic field~$\eta$ reads
\begin{align}
    D_\mu \eta &= \big( \partial_\mu - i g_3 G^A_\mu T^A - i g_2 W^I_\mu t^I  - i g_1 B_\mu \mathsf{y} \big)\eta \,,
\end{align}
where $g_{3,2,1}$ are the gauge couplings of $SU(3)_c$, $SU(2)_L$, and $U(1)_Y$, the corresponding generators are labeled $T^A$, $t^I$, and~$\mathsf{y}$, and the associated gauge fields are denoted $G_\mu^A$, $W_\mu^I$, and~$B_\mu$. } 
For simplicity, we assume that the $Z^\prime$ only couples to left-handed fermions, 
\begin{align}
    J_\mu &= g^{(q)}_{ij} \, \bar{q}_i \gamma_\mu q_j +g^{(l)}_{\alpha \beta} \, \bar{l}_\alpha \gamma_\mu l_\beta\,,
\end{align}
where $q$ and $l$ are SM quark and lepton doublets, with flavor indices denoted again by Latin and Greek symbols, respectively.~\footnote{We adopt the convention with diagonal down-quark Yukawas, so that the CKM matrix appears in the upper component of $q_i=[(V^\dagger\,u_{L})_i\,,d_{Li}]$.} Notice, in particular, that $Z^\prime$~couplings to fermions satisfy $\smash{g^{(q)}_{ij} = g^{(q)\,\ast}_{ji}}$ and $\smash{g^{(l)}_{\alpha\beta} = g^{(l)\,\ast}_{\beta\alpha}}$ due to Hermiticy. 

By integrating out the heavy~$Z^\prime$ from Eq.~\eqref{eq:Zprime} at tree level, we obtain the following effective Lagrangian with operators up to dimension~$d=8$,
\begin{align}
    \label{eq:Zprime-eft}
\mathcal{L}_\mathrm{eff}^{Z^\prime} \supset -\dfrac{J_\mu J^\mu}{2 m_{Z^\prime}^2} - \dfrac{1}{2 m_{Z^\prime}^4} \left(D_\mu J_\nu\right) \left(D^\mu J^\nu\right)\,,
\end{align}
where we have neglected the fermion masses. From this Lagrangian, we find the $d=6$ coefficient,
\begin{align}
    \dfrac{\big{[}\cC_{lq}^{(1)}\big{]}_{\alpha\beta ij}}{\Lambda^2} &= - \dfrac{g^{(l)}_{\alpha\beta}\, g^{(q)}_{ij}}{m_{Z^\prime}^2}\,, 
    \label{eq:Z'-matching-d6}
\end{align}
in addition to the $d=8$ ones,
\begin{align}
\dfrac{\big{[}\cC_{l^2q^2D^2}^{(1)}\big{]}_{2211}}{\Lambda^4} &= -\dfrac{g^{(l)}_{\alpha\beta}\, g^{(q)}_{ij}}{m_{Z^\prime}^4}\,.
\label{eq:Z'-matching-d8}
\end{align}
The corresponding operators are collected in Table~\ref{tab:SMEFT-ope}. The $d=8$ contributions given above correspond simply to the higher-order terms in the expansion of the $Z^\prime$~propagator in the $s$-channel.

\begin{table}[!t]
\renewcommand{\arraystretch}{2.2}
\centering
\begin{tabular}{c||c|c}
 Dim.\,\,\,\,& $\psi^4$ &   Operator\\ \hline\hline
 \multirow{2}{*}{$d=6$} & $\mathcal{O}_{lq}^{(1)}$ & $\big{(}\bar{l}_\alpha \gamma^\mu l_\beta\big{)}\big{(}\bar{q}_i \gamma_\mu q_j\big{)}$ \\ 
 & $\mathcal{O}_{lq}^{(3)}$ & $\big{(}\bar{l}_\alpha \gamma^\mu \tau^I l_\beta\big{)}\big{(}\bar{q}_i \gamma_\mu \tau^I q_j\big{)}$\\ \hline
 \multirow{4}{*}{$d=8$}&  $\mathcal{O}_{l^2 q^2 D^2}^{(1)}$  & $D^\nu (\bar l_\alpha \gamma^\mu l_\beta) D_\nu (\bar q_i \gamma_\mu q_j)$ \\
& $\mathcal{O}_{l^2 q^2 D^2}^{(2)}$ & $(\bar l_\alpha \gamma^\mu \overleftrightarrow{D}^\nu l_\beta) (\bar q_i \gamma_\mu \overleftrightarrow{D}_\nu q_j)$ \\
& $\mathcal{O}_{l^2 q^2 D^2}^{(3)}$ & $D^\nu (\bar l_\alpha \gamma^\mu \tau^I l_\beta) D_\nu (\bar q_i\gamma_\mu \tau^I q_j)$ \\
& $\mathcal{O}_{l^2 q^2 D^2}^{(4)}$ & $(\bar l_\alpha \gamma^\mu \overleftrightarrow{D}^{I\nu} l_\beta) (\bar q_i \gamma_\mu \overleftrightarrow{D}^I_\nu q_j)$
\end{tabular}
\vspace{0.2cm}
\caption{\small \sl 
Dimension $d=6$ and $d=8$ operators appearing in the matching to the concrete models in Sec.~\ref{sec:concrete} and which induce energy-enhanced contributions to the Drell-Yan cross section. Quark and lepton doublets are denoted by $q$ and~$l$, with flavor indices represented by Latin and Greek letters, respectively. The Pauli matrices are denoted by~$\tau^I$ with ${I\in\{1,2,3\}}$. We follow the conventions and notations of Ref.~\cite{Allwicher:2022gkm}.}
\label{tab:SMEFT-ope} 
\end{table}

\paragraph*{$U_1\sim(3,1,2/3)$\,:} 
The second scenario that we consider is a $U_1$~leptoquark with Lagrangian,~\footnote{Similarly to the $Z^\prime$ scenario, this model is non-renormalizable due to the presence of a massive vector boson.  It is possible to extend this model to generate the $U_1$~mass following, e.g.,~Ref.~\cite{DiLuzio:2017vat,Bordone:2018nbg}. }
\begin{align}
\label{eq:U1}
\begin{split}
 \hspace*{-.7em}   \cL_{U_1} \supset - \dfrac{1}{2} U_{1\;\!\mu\nu}^\dagger U_1^{\mu\nu} &+ m_{U}^2 \, U_1^{\mu\, \dagger} U_{1\,\mu} + (J_\mu^\dagger U_1^\mu + \mathrm{H.c.})\,,
\end{split}
\end{align}
where $U_{1\;\!\mu\nu}=D_\mu U_{1\nu}-D_\nu U_{1\mu}$ and, for simplicity, we consider only couplings to left-handed fermions
\begin{align}
    J_\mu^\dagger = x_{L}^{i\alpha}\,\bar q_i  \gamma_\mu l_\alpha\,. 
\end{align}
By integrating out the $U_1$~leptoquark at tree level, we find that the $d \leq 8$ Lagrangian is given by
\begin{align}
\label{eq:U1-eft}
\begin{split}
    \mathcal{L}_\mathrm{eff}^{U_1} \supset  - \dfrac{J_{\mu}^\dagger J^\mu}{ m_{U_1}^2} &- \dfrac{1}{m_{U_1}^4} \left(D_\mu J_\nu\right)^\dagger \left(D^\mu J^\nu\right)\\ 
    &+ \dfrac{1}{m_{U_1}^4} \left(D_\mu J_\nu\right)^\dagger \left(D^\nu J^\mu\right)\,.
\end{split}
\end{align}
We apply $SU(2)_L$ and Dirac Fierz relations to reduce the above Lagrangian to the Warsaw basis of $d=6$ operators, where the only non-vanishing coefficients are
\begin{align}
    \dfrac{\big{[}\cC_{lq}^{(1)}\big{]}_{\alpha\beta ij}}{\Lambda^2} &= \dfrac{\big{[}\cC_{lq}^{(3)}\big{]}_{\alpha\beta ij}}{\Lambda^2} = -\dfrac{x_{L}^{i \beta} x_L^{j \alpha\,\ast}}{2 m_{U_1}^2}\,.
\end{align}
Moreover, we find that several operators appear at~$d=8$, including those that are not energy enhanced and thus neglected here. The only operators that induce the maximal energy scaling of the amplitudes (i.e.,~$\propto E^4/\Lambda^4$) contain two additional derivatives with respect to the $d=6$ terms. Their coefficients read
\begin{align}
\dfrac{\big{[}\cC_{l^2q^2D^2}^{(n)}\big{]}_{\alpha\beta ij}}{\Lambda^4} = (-1)^{n+1}\, \frac{ x_{L}^{i \beta} x_L^{j \alpha\,\ast}}{4 m_{U_1}^4}\,,
\end{align}
for $n\in \lbrace 1,2,3,4\rbrace$, with flavor indices denoted as above, cf.~Table~\ref{tab:SMEFT-ope}. Note, in particular, that the same combinations of leptoquark couplings appear in the numerators of the $d=6$ and $d=8$ coefficients.

\section{EFT vs.~concrete models}
\label{sec:eft-vs-models}

In this Section, we compare the full predictions from the concrete models introduced in Sec.~\ref{sec:concrete} with the ones obtained employing the EFT Lagrangian, which has been matched to these models at a given order in the $1/\Lambda$ expansion. This comparison will first be made at the parton level using the analytical expressions in Sec.~\ref{ssec:parton-formulas}, which will then be convoluted with the PDFs for a numerical comparison in Sec.~\ref{ssec:numerical-comparison}.

\subsection{Partonic description}
\label{ssec:parton-formulas}

\paragraph*{$Z'\sim ({\bf 1}, {\bf 1}, 0)$} Firstly, we consider the $Z^\prime$~model introduced in Eq.~\eqref{eq:Zprime} and we compute the partonic cross section for $q_i \bar{q}_j \to \ell_\alpha^- \ell_\alpha^+$ by using the expressions of Sec.~\ref{sec:drell-yan} for both the EFT, which is matched to the $Z^\prime$~model, as well as the full-model calculation. For definiteness, we consider down-quark transitions (i.e., $q=d$), but our expressions can be extended \emph{mutatis mutandis} to $q=u$. The full cross section normalized by the SM one reads~\footnote{This expression can be easily extended to the lepton flavor violating case by removing the interference term.}
\begin{align}
\label{eq:x-section-Zp}
\dfrac{\hat{\sigma}^{Z^\prime}}{\hat{\sigma}^{\mathrm{SM}}} = 1+ 2 a_L \,\delta_{ij}\,\mathrm{Re}\left[\dfrac{g_{ij}^{(q)}g_{\alpha\alpha}^{(l)}}{1-x_V^{-1}}\right] + b_L\,\left|\dfrac{g_{ij}^{(q)}g_{\alpha\alpha}^{(l)}}{1-x_V^{-1}}\right|^2\,,
\end{align}
where $x_V \equiv \hat{s}/\Omega_{Z^\prime}$, $\Omega_{Z^\prime}=m_{Z^\prime}^2-i m_{Z^\prime}\Gamma_{Z^\prime}$, and $\Gamma_{Z^\prime}$ is the $Z^\prime$~width, and we have factored out the following pre-factors
\begin{align}
\label{eq:a-b-factors}
a_L &\equiv\dfrac{\mathcal{F}_{LL}^{\mathrm{SM}}\,v^2/\hat{s}}{\sum_{X,Y}|\mathcal{F}_{XY}^\mathrm{SM}|^2} \,,\qquad 
b_L \equiv \dfrac{v^4/\hat{s}^2}{\sum_{X,Y}|\mathcal{F}_{XY}^\mathrm{SM}|^2}\,,
\end{align}
as they become constant in the limit $v^2/\hat{s}\to 0$ [cf.~Eq.~\eqref{eq:F-SM}]. Assuming the couplings to be real and neglecting the $Z^\prime$~width, the fractions in Eq.~\eqref{eq:x-section-Zp} are a geometric series in $\hat{s}/m_{Z^\prime}^2$ and its square, which can be expanded for $x_V < 1$ (i.e., $\hat{s} < m_{Z^\prime}^2$) 
\begin{align}
\label{eq:Taylor-Zp}
\hspace{-0.4em}
\dfrac{1}{1-x_V^{-1}} = - \sum_{n=1}^\infty x_V^n \,,
\quad
\dfrac{1}{\left(1-x_V^{-1}\right)^2} = \sum_{n=1}^\infty n\,x_V^{n+1} \,,
\end{align}
leading to 
\begin{align}
\label{eq:x-section-Zp-expanded}
\dfrac{\hat{\sigma}^{Z^\prime}}{\hat{\sigma}^{\mathrm{SM}}}& = 1 - 2\,a_{L}\, \delta_{ij} \,g_{ij}^{(q)}g_{\alpha\beta}^{(\ell)}\dfrac{\hat{s}}{m_{Z^\prime}^2} \\
&+\Big{[}b_{L}\,|g_{ij}^{(q)}g_{\alpha\beta}^{(\ell)}|^2-2a_{L}\,\delta_{ij}\,g_{ij}^{(q)}g_{\alpha\beta}^{(\ell)}\Big{]} \dfrac{\hat{s}^2}{m_{Z^\prime}^4} + \mathcal{O}(m_{Z^\prime}^{-6})\,,\nonumber
\end{align}
from which the energy enhancement of the cross section becomes clear. Most importantly, the expansion of the $Z^\prime$~propagator in Eq.~\eqref{eq:x-section-Zp} is slowly convergent as $\hat{s}$ approaches~$m_{Z^\prime}^2$, since all the terms in the geometric series contribute with the same sign. This is expected from the resonant nature of this process.

\paragraph*{$U_1\sim(3,1,2/3)$} We turn now our attention to the $U_1$~leptoquark model defined in Eq.~\eqref{eq:U1}. The ratio of the full $d_i\bar{d}_j \to \ell_\alpha^-\ell_\alpha^+$ cross section with respect to the SM one can be written in a similar way after integration over $\hat{t}\in (-\hat{s},0)$. Neglecting the leptoquark width, we can write~\cite{Jaffredo:2021ymt,Davidson:2013fxa,Davidson:2014lsa,Eboli:1987vb}
\begin{align}
\label{eq:x-section-U1}
\dfrac{\hat{\sigma}^{U_1}}{\hat{\sigma}^{\mathrm{SM}}} = 1 &+ 2a_L\,\delta_{ij}\,\varphi_1(x_V)\,\mathrm{Re}\big[x_L^{i\alpha}x_L^{j\alpha\,\ast}\big]\nonumber\\* 
&\qquad\qquad + b_L\, \varphi_2(x_{V})\,\big|x_L^{i\alpha}x_L^{j\alpha\,\ast}\big|^2\,,
\end{align}
where $x_V\equiv\hat{s}/m_{U_1}^2$, and the same pre-factors defined in Eq.~\eqref{eq:a-b-factors} are factored out. The phase-space functions $\varphi_k$ ($k=1,2$) now involve the integral over the $t$-channel leptoquark propagator,
\begin{align}
\label{eq:phi-int}
\varphi_k(x_V) & \equiv \int_{-1}^{0}  \mathrm{d}\omega\, \dfrac{3(1+\omega)^2}{(\omega-x_V^{-1})^k}\,,
\end{align}
where $\omega \equiv \hat{t}/\hat{s}$. The full expressions for these integrals are given in Appendix~\ref{app:lq-formulas}, while here we express them as power series for $x_V < 1$,
\begin{align}
\label{eq:phi-series}
\varphi_1(x_V) &= 6 \sum_{n=1}^{\infty} \dfrac{(-1)^n \,x_V^n}{n (n+1) (n+2)}\,, \\*[0.4em]
\varphi_2(x_V) &= 6 \sum_{n=2}^{\infty} \dfrac{(-1)^n \,x_V^n}{n (n+1)}\,. \nonumber
\end{align}
Notice, in particular, that this series converges faster than the $Z^\prime$-model one in Eq.~\eqref{eq:Taylor-Zp}, due to the alternating signs of the sub-leading corrections, which are attributed to higher-dimensional operators in the EFT formalism. Similarly to the above calculation, we assume the couplings to be real and we keep the first terms of the $x_V$~expansion,
\begin{align}
\label{eq:x-section-U1-expanded}
\dfrac{\hat{\sigma}^{U_1}}{\hat{\sigma}^{\mathrm{SM}}} &= 1 - 2\,a_{L}\, \delta_{ij} \, \mathrm{Re}\big[x_L^{i\alpha}x_L^{j\alpha\,\ast}\big]\dfrac{\hat{s}}{m_{U_1}^2} \\
&+\!\Big{[}b_{L}\,\big|x_L^{i\alpha}x_L^{j\alpha\,\ast}\big|^2 \!+\frac{a_{L}}{2}\,\delta_{ij}\,x_L^{i\alpha}x_L^{j\alpha\,\ast}\Big{]} \dfrac{\hat{s}^2}{m_{U_1}^4} + \mathcal{O}(m_{U_1}^{-6})\,,\nonumber
\end{align}
Although we focused on a specific leptoquark scenario, we stress that the above conclusions are valid for any tree-level mediator in $u$- or $t$-channels.

\begin{figure*}[!t]
    \includegraphics[width=0.32\textwidth]{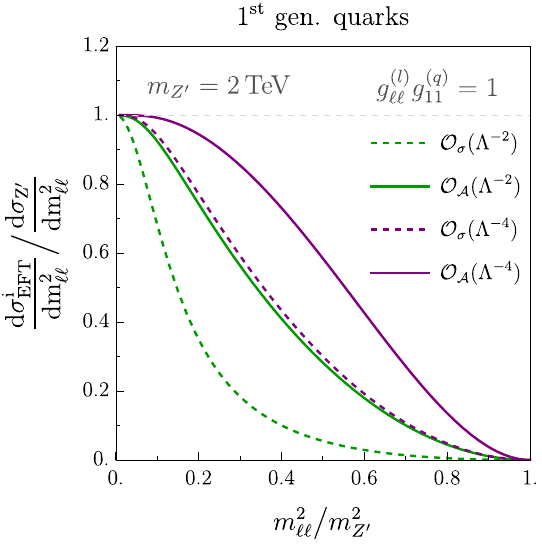}     
    \includegraphics[width=0.32\textwidth]{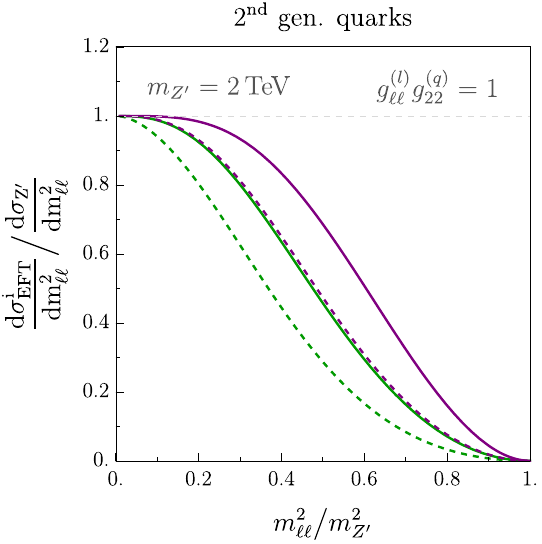}
    \includegraphics[width=0.32\textwidth]{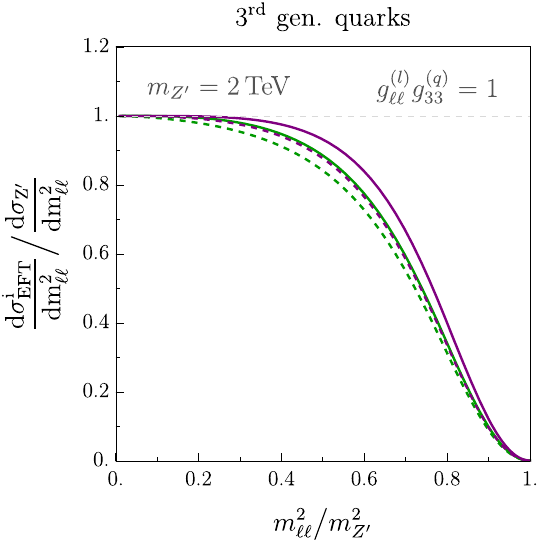} \\
    \includegraphics[width=0.32\textwidth]{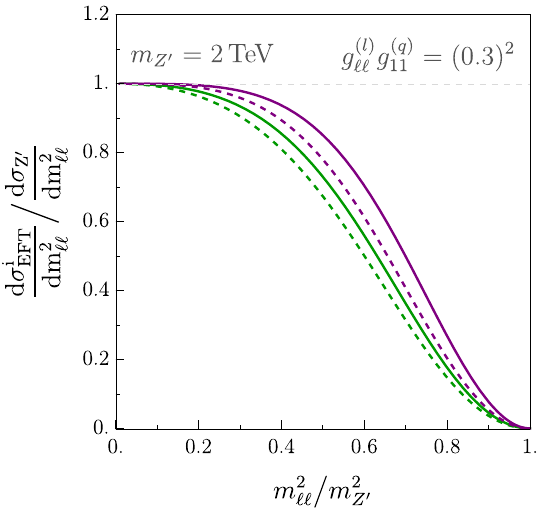}
    \includegraphics[width=0.32\textwidth]{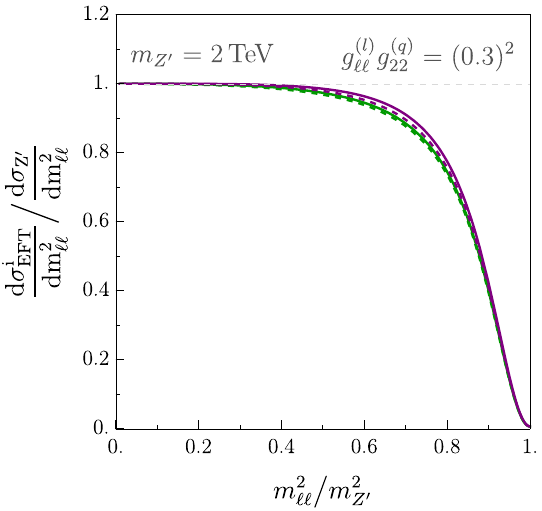}
    \includegraphics[width=0.32\textwidth]{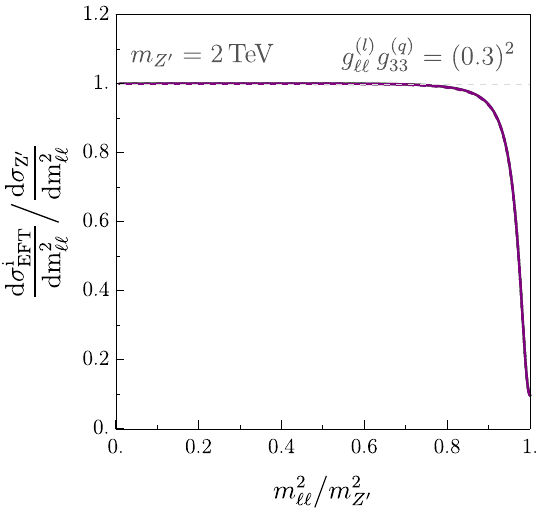}
    \caption{\small \sl The ratio of the differential cross section in the EFT approach ($\mathrm{d}\hat{\sigma}_{\mathrm{EFT}}/\mathrm{d}m_{\ell\ell}^2$) and the $Z^\prime\sim (\mathbf{1},\mathbf{1},0)$ model ($\mathrm{d}\hat{\sigma}_{Z^\prime}/\mathrm{d}m_{\ell\ell}^2$) are plotted as a function of $\smash{m_{\ell\ell}^2/m_{Z^\prime}^2}$, where $m_{\ell\ell}=\sqrt{\hat{s}}$ is the invariant mass of the dilepton system. The mediator mass is fixed to $m_{Z^\prime}=2~\mathrm{TeV}$ and the couplings to $\smash{g^{(l)}_{\ell\ell}\,g^{(q)}_{ii}=1}$ (upper row) and $\smash{g^{(l)}_{\ell\ell}\,g^{(q)}_{ii}=(0.3)^2}$ (lower row), for fixed lepton flavors and different quark flavors in the columns of this plot. The EFT cross section is computed at different orders in the EFT expansion, with contributions up to interference of $d=6$ terms with the SM (dashed green), squared $d=6$ terms (solid green), interference of $d=8$ terms with the SM (dashed purple), and squared $d=8$ terms (solid purple).
}
    \label{fig:diffcrosssection-Zp}
\end{figure*}

\begin{figure*}[!t]
   \includegraphics[width=0.32\textwidth]{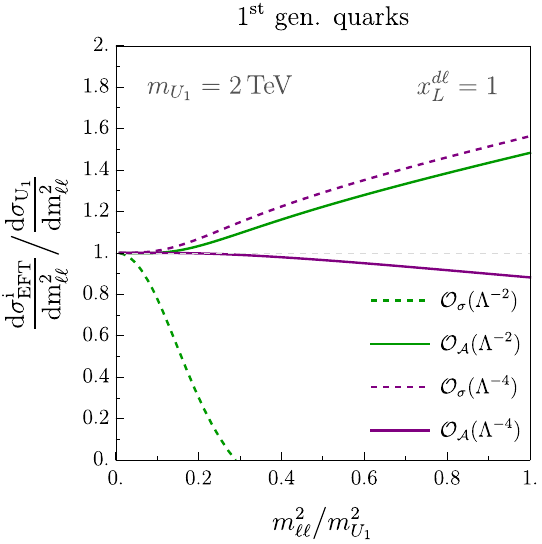}
    \includegraphics[width=0.32\textwidth]{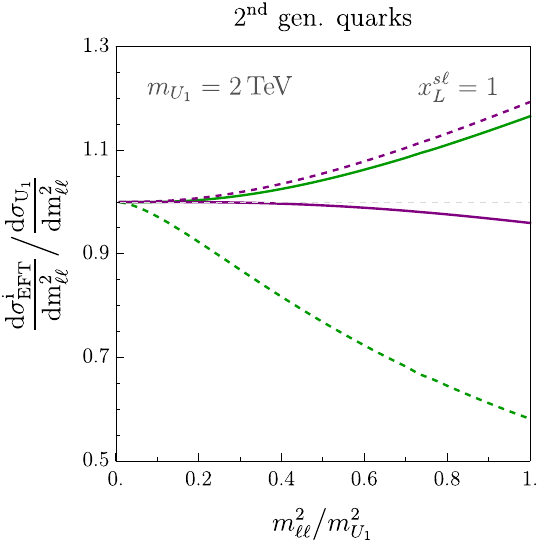}
    \includegraphics[width=0.32\textwidth]{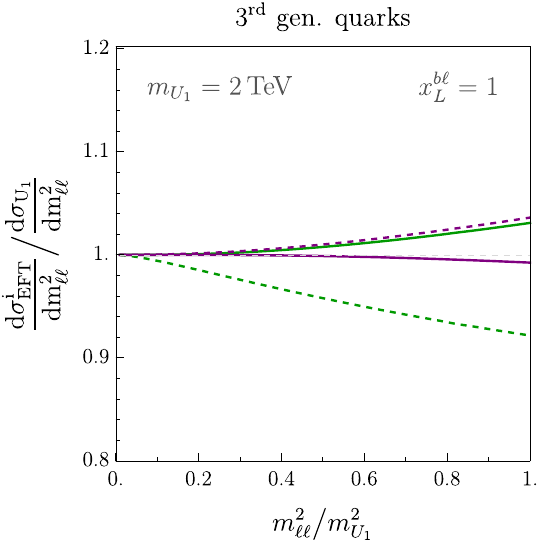} \\
    \includegraphics[width=0.32\textwidth]{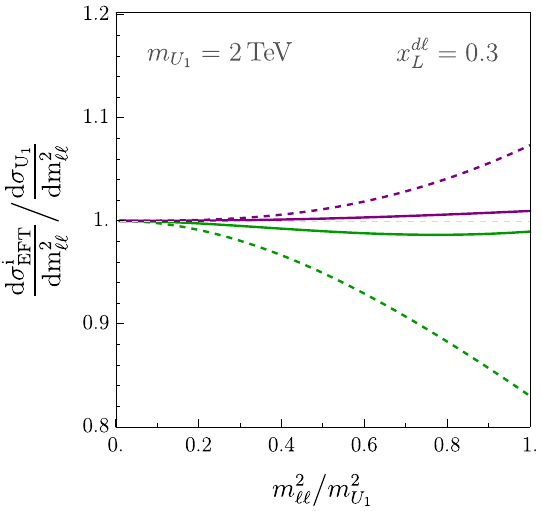}
    \includegraphics[width=0.32\textwidth]{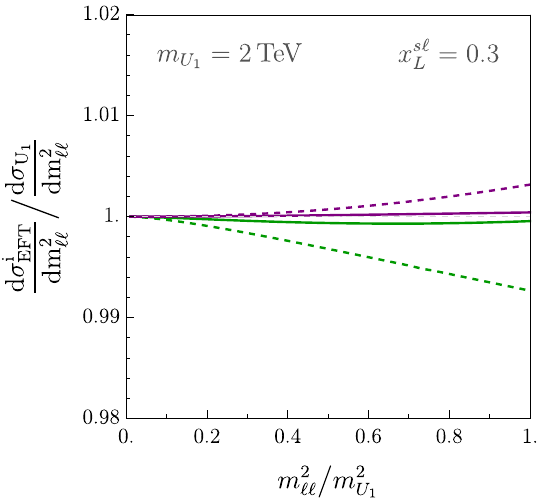}
    \includegraphics[width=0.32\textwidth]{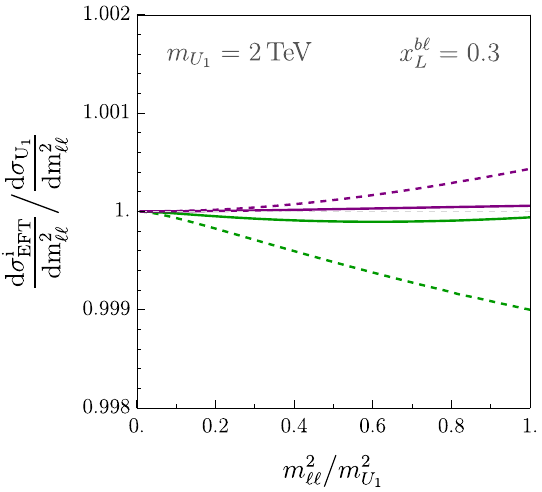}
    \caption{\small \sl Comparison between EFT and full-model cross section for the $U_1\sim (\mathbf{3},\mathbf{1},2/3)$ leptoquark with fixed mass $m_{U_1}=2~\mathrm{TeV}$ and couplings $x_{L}^{i\alpha}=1$ (upper row) and $x_{L}^{i\alpha}=0.3$ (lower row). See caption of Fig.~\ref{fig:diffcrosssection-Zp} for details.}
    \label{fig:diffcrosssection-U1}
\end{figure*}

\subsection{Cross sections and quark flavor dependence}
\label{ssec:numerical-comparison}

Next, we consider the effects on the previous conclusions from convoluting the partonic cross sections with the different quark PDFs. 
The main expected impact is a difference between scenarios with mediators coupled to light or heavy fermions, for which the EFT convergence can be studied numerically.

In Fig.~\ref{fig:diffcrosssection-Zp}, we compare the EFT cross sections to the predictions of the full $Z^\prime$~model, as a function of the ratio $m_{\ell\ell}^2/m_{Z^\prime}^2$, for a fixed mass $m_{Z^\prime}=2~\mathrm{TeV}$ and fixed lepton flavor $\ell$, where $m_{\ell\ell}=\smash{\sqrt{\hat{s}}}$ is the dilepton invariant mass.~\footnote{The full cross sections have been computed at tree level using the \texttt{HighPT} package, not relying on the expansions shown in Eqs.~\eqref{eq:x-section-Zp-expanded} and~\eqref{eq:x-section-U1-expanded}.}
Specifically, the ratio of the differential EFT cross section, computed up to different truncation orders, to that obtained in the full model is plotted.
The dashed green line is determined only taking into account the interference of $d=6$ EFT operators with the SM amplitude, while the solid green, dashed purple, and solid purple lines sequentially add contributions from the square of $d=6$ terms, the interference of $d=8$ operators with the SM, and the square of the $d=8$ terms, respectively.
Notice that the dashed lines correspond to truncating the EFT series on cross-section level at $\cO_{\!\sigma}(\Lambda^{-2})$ and $\cO_{\!\sigma}(\Lambda^{-4})$ for the green and purple lines, respectively.
The solid lines, on the other hand, represent a truncation on amplitude level at $\cO_{\!\cA}(\Lambda^{-2})$ and $\cO_{\!\cA}(\Lambda^{-4})$ again for the green and purple lines, respectively. 
Here, we have introduced the notation $\cO_{\!\sigma}(\Lambda^{-n})$ and $\cO_{\!\cA}(\Lambda^{-n})$ to indicate a truncation of the EFT series on the cross-section and amplitude level, respectively.
A~further discussion of the two truncation approaches is presented in Sec.~\ref{sec:truncation}.
To obtain the EFT cross section, the matching conditions in Eqs.~\eqref{eq:Z'-matching-d6} and~\eqref{eq:Z'-matching-d8} are employed.
The light gray dashed line indicates the cross section of the full model.
The couplings are fixed to $\smash{g^{(\ell)} g^{(q)} = 1}$ (upper row) and $\smash{g^{(\ell)} g^{(q)} = (0.3)^2}$ (lower row), with flavor-diagonal couplings to quarks and leptons. 
The flavor of the quarks to which the $Z^\prime$ is coupled is varied from the first (left column) to the second (center column) and third (right column) generation.
A similar comparison is made in Fig.~\ref{fig:diffcrosssection-U1} for the $U_1$~leptoquark model, again considering two fixed values of the couplings~$x_L^{q\ell}$ and different quark flavors.

\subsection{Discussion}

Several comments can be made by comparing the EFT and the full model predictions in Fig.~\ref{fig:diffcrosssection-Zp} and~\ref{fig:diffcrosssection-U1}:
\begin{itemize}
    \item[{\it i})]
    In all cases, we retrieve the EFT regime for $m_{\ell\ell}^2/m_{V}^2 \to 0$, that is, at low center-of-mass energies relative to the mediator mass~$m_V$, the cross section of the full model is well approximated by the interference of $d=6$~EFT amplitude with the SM (dashed green line).
    \item[{\it ii})] 
    As soon as $m_{\ell\ell}^2/m_{V}^2$ increases, it is necessary to include contributions beyond the $d=6$ interference with the SM to correctly describe the full model cross sections. 
    The square of $d=6$ operators (solid green), which we denote by $(d=6)^2$, are necessary in most cases to obtain accurate predictions, whereas $d=8$ terms can improve the accuracy of the predictions. 
    However, notice that including only the linear $d=8$ terms interfering with the SM (dashed purple) offers only marginally better limits, whereas a significant improvement can be observed when including the $(d=8)^2$ contributions (solid purple).
    \item[{\it iii})]
    In all considered cases, the EFT description appears to converge better toward the full model when truncating the EFT series on amplitude level~$\cO_{\!\cA}(\Lambda^{-n})$ (solid lines), rather than on cross-section level~$\cO_{\!\sigma}(\Lambda^{-n})$ (dashed lines), i.e., when including New-Physics squared contributions at both $d=6$ and $d=8$ rather than just the corresponding interference terms with the SM.
    For a more detailed discussion, see Sec.~\ref{sec:truncation}.
    \item[{\it iv})] 
    The EFT description provides a more accurate description of these processes for much larger values of $m_{\ell\ell}^2/m_{V}^2$ for the leptoquark model than for the~$Z^\prime$~one, as expected from the resonant nature of the latter scenario. 
    This can be understood in terms of the power series for the cross sections, which has terms with alternating and same signs, respectively [cf.~Eqs.~\eqref{eq:phi-series} and~\eqref{eq:Taylor-Zp}].
    Notably, when including $(d=8)^2$ contributions (solid purple) for the leptoquark model, the discrepancy of the EFT approximation and the full model is below the $\lesssim 10\,\%$~level even for $m_{\ell\ell}^2/m_V^2 \sim 1$ and $\cO(1)$~couplings.
    In contrast, we find that the EFT series is converging at a significantly slower rate for the $Z^\prime$~scenario, when including higher-order terms, which only gradually improve the accuracy of the EFT approximation.
    \item[{\it v})] 
    Finally, the EFT convergence depends on the size of the New-Physics couplings convoluted with the quark PDFs. 
    The EFT terms converge better for small couplings and/or small PDFs (i.e.,~heavy flavors) since the New-Physics contributions to the hadronic cross section are smaller in this case. 
\end{itemize}

Note also that we neglect Renormalization Group~(RG) evolution effects between the different energy scales probed by the Drell-Yan data.
Heavy new physics scenarios are mostly sensitive to the high-$p_T$ tails of these distributions where RG effects are small.
Moreover, the RG effects are similar between a UV theory and its corresponding low-energy EFT. 
Therefore, when the ratio of the cross sections is taken, we expect these effects to largely cancel out and not affect the study of the EFT validity.
However, for a broader phenomenological analysis (which is beyond the scope of the present work), these effects should be considered~\cite{Bartocci:2024fmm}.

\subsection{EFT series truncation: amplitude vs.~cross-section level}
\label{sec:truncation}

In this Section, we briefly elaborate on the different approaches for truncating the EFT series that have been used in Figs.~\ref{fig:diffcrosssection-Zp} and \ref{fig:diffcrosssection-U1}.
In particular, we will contrast the truncation on cross-section level, i.e., for physical observables, denoted by~$\cO_{\!\sigma}(\Lambda^{-n})$ in our notation, against the truncation on amplitude level represented by~$\cO_{\!\cA}(\Lambda^{-n})$.
Schematically, we can express the amplitude~$\cA$ and partonic cross section~$\hat{\sigma}$ in terms of the individual EFT-order contributions as
\begin{widetext}
\begin{align} \label{eq:cross-section_series}
    {\hat{\sigma}} &\propto \left\lvert \cA_{\mathrm{SM}} + \frac{\cA_6}{\Lambda^2} + \frac{\cA_{6 \times 6} + \cA_8}{\Lambda^4} + \cO_{\!\cA}(\Lambda^{-6}) \right\rvert^2
   \\*
    &= 
    \underbrace{\underbrace{\left\lvert \cA_\mathrm{SM} \right\rvert^2 
    + \frac{2 \mathrm{Re} \left( \cA_\mathrm{SM}^\ast \cA_6 \right)}{\Lambda^2}
    + \frac{\left\lvert \cA_\mathrm{6} \right\rvert^2}{\Lambda^4}}_{\cO_{\!\cA}(\Lambda^{-2})} 
    + \frac{2 \mathrm{Re} \left( \cA_\mathrm{SM}^\ast \cA_{6 \times 6} + \cA_\mathrm{SM}^\ast \cA_8 \right)}{\Lambda^4} 
    + \frac{2 \mathrm{Re} \left( \cA_6^\ast \cA_{6 \times 6} + \cA_6^\ast \cA_8 \right)}{\Lambda^6} 
    + \frac{\left\lvert \cA_\mathrm{6 \times 6} \right\rvert^2 + \left\lvert \cA_\mathrm{8} \right\rvert^2}{\Lambda^8}}_{\cO_{\!\cA}(\Lambda^{-4})} 
    + \ldots
    \nonumber \,,
\end{align}
\end{widetext}
where $\cA_{\mathrm{SM}},\,\cA_{6},\,\cA_{6 \times 6}$, and~$\cA_{8}$ represent the amplitudes of the SM, the EFT with a single $d=6$ operator insertion, with a double $d=6$ operator insertion, and a single $d=8$ operator insertion, respectively. 
The EFT power counting is made explicit here by factoring out the appropriate powers of the EFT cutoff scale~$\Lambda$. 
The ellipsis indicate terms of order~$\cO_{\!\cA}(\Lambda^{-6})$ in the EFT expansion of the amplitude.
For low-energy measurements (e.g.,~in the flavor sector) considering only the linear $d=6$ interference terms of~$\cO_{\!\sigma}(\Lambda^{-2})$ suffices and the $(d=6)^2$ contributions of~$\cO_{\!\cA}(\Lambda^{-2})$ can be neglected due to the large scale suppression.
Investigating the EFT truncation for high-energy observables, such as the Drell-Yan tails, is more subtle.
The shapes of distributions for high-$p_T$ observables are determined by the high-energy properties of the corresponding amplitudes, which in turn are governed by their analytical structure and the principle of unitarity~\cite{Adams:2006sv}, even though the latter is lost for the EFT at very high energies $E>\Lambda$ above the cutoff.~\footnote{Notice, in particular, that the Wilson coefficients of $d=8$ operators of the considered class $\cO_{\psi^4 D^2}$ are also subject to positivity constraints~\cite{Chala:2023xjy} (see also Ref.~\cite{Remmen:2019cyz}).}
Here, however, we consider energies where the unitarity is not yet lost. 
When truncating the EFT series on amplitude level~$\cO_{\!\cA}(\Lambda^{-n})$ for a given $n\in\mathbb{N}$, these properties are retained below the cutoff $E<\Lambda$.
However, when truncating on cross-section level~$\cO_{\!\sigma}(\Lambda^{-n})$ these properties are upset since this does not correspond to a consistent truncation of the amplitude.
In the latter case, even negative values for cross sections are, in principle, possible.

Furthermore, we notice that the EFT series is unphysical. 
This can be understood through the LSZ formula~\cite{Lehmann:1954rq} which permits the freedom to perform field redefinitions for a theory without changing its $S$~matrix, that is, physical observables. 
However, in the case of an EFT, these field redefinitions generally shift contributions between different orders in the EFT expansions, rendering the expansion unphysical.
Therefore, it appears that requiring a consistent EFT truncation for physical observables is unwarranted.
Instead, it is useful to adopt a top-down perspective to determine a consistent EFT truncation prescription for high-energy observables.
While the EFT is useful for studying experimental data in a model-independent framework, it should ultimately always be matched onto concrete underlying UV~models of interest.
When this is done, the EFT is constructed (or in other words, its coefficients are adjusted) in such a way that the resulting EFT amplitudes are mimicking the corresponding UV amplitude to a given accuracy in the EFT power counting, see e.g.~Ref.~\cite{DeAngelis:2023bmd}.~\footnote{
This is the case for both \emph{off-shell} and \emph{on-shell matching}, where the corresponding EFT amplitudes are equated with the ones of the full UV~model in order to determine the matching conditions.}
This naturally suggests that also in a bottom-up scenarios, the EFT series should be truncated on amplitude rather than cross-section level, at least if the bottom-up analysis is intended to be linked to realistic UV scenarios at a later stage.~\footnote{In a pure bottom-up setup, where there is no intention of linking to UV theories, this strategy is, of course, less compelling.}
Only the former truncation method guarantees that the EFT properly reproduces the behavior of UV models.
We therefore suggest truncating the EFT series on amplitude level~$\cO_{\!\cA}(\Lambda^{-n})$ for some $n\in\mathbb{N}$.
As explicitly verified for Drell-Yan processes in Figs.~\ref{fig:diffcrosssection-Zp} and~\ref{fig:diffcrosssection-U1} with $n=2$ and $4$ the $\cO_{\!\cA}(\Lambda^{-n})$~truncations (solid lines) provide better convergence toward the full model prediction than the cross-section truncation~$\cO_{\!\sigma}(\Lambda^{-n})$ (dashed lines).
According to this prescription, the leading order~$\cO_{\!\cA}(\Lambda^{-2})$ EFT contributions contain both interference of $d=6$ EFT operators with the SM and the $(d=6)^2$ contributions to the cross section, which can also be seen in Eq.~\eqref{eq:cross-section_series}. 
As discussed in Ref.~\cite{Brivio:2022pyi}, this provides a well-defined, unambiguous, and gauge-invariant setup. 
At the first subleading order~$\cO_{\!\cA}(\Lambda^{-4})$ interference of $d=8$ operators with the SM, $(d=8)^2$ contributions, as well as the interference of $(d=6)$ with $(d=8)$ operators are included. 
In addition to dimension-eight operators contributing to the amplitude, the corresponding contributions from double insertion of two $d=6$ operators into the same amplitude are taken into account at this order as well, cf.~Eq.~\eqref{eq:cross-section_series}, although this is not relevant for the present discussion of Drell-Yan tails at tree level.

\section{Clipped limits}
\label{sec:clipping}

\begin{figure*}[!p]
\includegraphics[width=0.95\textwidth]{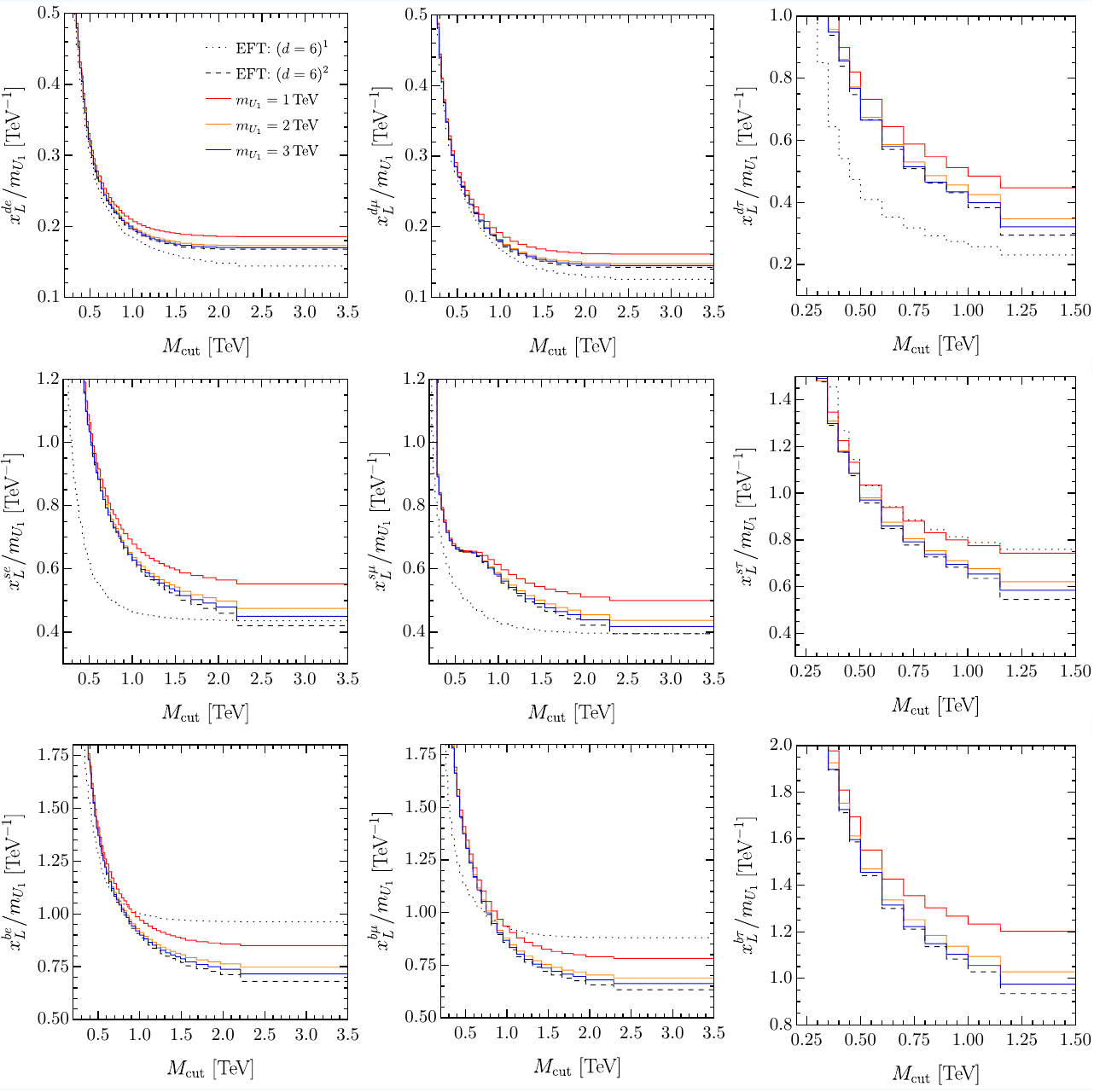}
    \caption{\small \sl  Expected limits on the $U_1$~leptoquark coupling defined in Eq.~\eqref{eq:U1} from $pp\to \ell\ell$ data for $\ell=e,\mu$ (left and center column)~\cite{CMS:2021ctt} and $\ell=\tau$ (right column)~\cite{ATLAS:2020zms} are plotted against the clipping variable~$M_{\mathrm{cut}}$, with bins above~$M_{\mathrm{cut}}$ being discarded. For $\ell=e,\mu$ we remove events with $m_{\ell\ell}^\mathrm{rec}> M_{\mathrm{cut}}$, while we discard events with $m_T^\mathrm{tot}>M_{\mathrm{cut}}$ for $\ell=\tau$. 
    The top, center, and bottom row show the case of couplings to first-, second-, and third-generation quarks, respectively.
    The dotted and dashed lines correspond to the EFT constraints matched to the leptoquark model with contributions up to $(d=6)$ interference and $(d=6)^2$ terms, i.e., $\mathcal{O}_{{\sigma}}(\Lambda^{-2})$ and $\mathcal{O}_{\mathcal{A}}(\Lambda^{-2})$  truncations, respectively. 
    The leptoquark mass is fixed to different values to illustrate the EFT convergence, namely $m_{U_1}=1~\mathrm{TeV}$ (red), $m_{U_1}=2~\mathrm{TeV}$ (orange) and $m_{U_1}=3~\mathrm{TeV}$ (blue).}
    \label{fig:U1clipping}
\end{figure*}

Notice that the flavor of leptons is indifferent for the predictions in Figs.~\ref{fig:diffcrosssection-Zp} and~\ref{fig:diffcrosssection-U1}. 
However, it affects the interpretation of experimental measurements and, as a consequence, the robustness of the EFT description, since the data sets collected for $\ell=e,\mu$ include events at higher energies (i.e.~with larger~$m_{\ell\ell}$ values) compared to the data for $\ell=\tau$, and since the latter is also affected by uncertainties from the reconstruction of the hadronic $\tau$~decays.
Therefore, higher energy scales are probed for light leptons, typically in the $\mathcal{O}(10~\mathrm{TeV})$ range, whereas $\mathcal{O}(1~\mathrm{TeV})$ scales are accessible for $\tau$~leptons, as shown e.g.~in Ref.~\cite{Allwicher:2022gkm}. 

The comparison between our EFT and full model predictions with the LHC data is made for the $U_1$~leptoquark in Fig.~\ref{fig:U1clipping}. 
The expected upper limits on the coupling-over-mass ratio $|x_{L}^{i\alpha}|/m_{U_1}$ are shown as a function of the \emph{clipping} variable~$M_{\mathrm{cut}}$.
More precisely, the constraints are determined only taking into account the experimental data below the threshold scale given by~$M_{\mathrm{cut}}$. 
For light leptons $\ell = e,\mu$, the clipping is performed for the reconstructed invariant mass of the dilepton system~$m_{\ell\ell}^\mathrm{rec}$, i.e., considering only data with $m_{\ell\ell}^\mathrm{rec} < M_{\mathrm{cut}}$, while for third-generation leptons $\ell=\tau$, we clip the total traverse mass~$m_T^\mathrm{tot}$ of the ditau system, which is used as experimental observable in this case~\cite{ATLAS:2020zms}. 

We have also verified that clipping the New-Physics signal prediction on the partonic center-of-mass energy~$\sqrt{\hat{s}}=m_{\ell\ell}$, rather than clipping on experimental observable level, yields comparable results. For light leptons $\ell=e,\mu$ we find good agreement, since we have $m_{\ell\ell}^\mathrm{rec} \simeq m_{\ell\ell}$. For $\tau$~leptons, we find some differences due to the mismatch between $m_T^\mathrm{tot}$ and~$m_{\ell\ell}$. We actually have $m_T^\mathrm{tot} \leq m_{\ell\ell}$ and thus events with higher center-of-mass energies can migrate down to lower values of the observed energy variable~$m_T^\mathrm{tot}$.
It is hence not obvious how a consistent signal clipping prescription can be introduced in this case.
For a discussion of the problems when clipping the signal rather than the data see e.g.~Ref.~\cite{Falkowski:2016cxu,Brivio:2022pyi}. 
However, the patterns of the EFT convergence observed in the present example are similar for clipping data and signal.

For definiteness, we consider the leptoquark coupled to first-, second-, and third-generation quarks (top, center, and bottom row, respectively), with different lepton flavors that are constrained by LHC data on $pp\to\ell\ell$ with $\ell=e,\mu$~\cite{CMS:2021ctt} (left and center column) and $\ell=\tau$~\cite{ATLAS:2020zms} (right column).
These searches have been reinterpreted in the {\tt HighPT} package~\cite{Allwicher:2022mcg}, which we employ to determine the limits presented here.~\footnote{For the $pp\to\tau\tau$ channel, we use an updated version of {\tt HighPT}~\cite{Allwicher:2022mcg} in which the selections cuts are improved to better reproduce the experimental search, which is highly sensitive to the $p_T$~cut for the leading $\tau_h$~jet. These improved constraints will soon be publicly available in a forthcoming updated version of \texttt{HighPT}.} 
The dotted and dashed lines represent the $2\sigma$~constraints obtained using the EFT approach and taking into account only the $d=6$ interference with the SM, or additionally also the $(d=6)^2$ contribution.
We also show the corresponding limits determined in the full model with leptoquark masses~$m_{U_1}$ of 1,~2, and 3~TeV in red, yellow, and blue, respectively.

First of all, we notice that the constraints obtained in the full model converge well for all leptoquark masses toward the EFT limits, computed considering $(d=6)^2$ contributions (dashed line), in the limit $M_\mathrm{cut} \to 0$, i.e., when only considering low-energy data.
On the contrary, the EFT limits, calculated taking into account only the $d=6$ interference with the SM (dotted line), do not converge in this limit. 
This can be understood by realizing that the larger the couplings, the more important the $(d=6)^2$ terms become relative to the linear $d=6$ interference contribution.
However, when clipping at low values of~$M_\mathrm{cut}$, most of the data is discarded and we are hence obtaining weaker constraints in the large coupling regime.~\footnote{%
This is the case at least when coupling to second- and third-generation quarks. 
For first-generation quarks (top row in Fig.~\ref{fig:U1clipping}) the valence-quark PDF enhancement still allows to probe relatively small values for the coupling, where the $d=6$ interference term dominates and thus converges to the full model.}
Thus, the leading contribution in the $M_\mathrm{cut} \to 0$ limit is given by the $(d=6)^2$ terms (at least for sea quarks) and the linear $d=6$ interference terms cannot approximate the full model well.
Also for larger values of~$M_\mathrm{cut}$, we find that the models always converge toward the $(d=6)^2$ EFT limits, when increasing the leptoquark mass.
Therefore, we conclude with this concrete example that using only the linear $d=6$ interference terms does not provide an accurate description for the Drell-Yan tails and that $(d=6)^2$ contributions should always be included.

In addition, we find in Fig.~\ref{fig:U1clipping}, as expected, that the heavier the leptoquark mass, the better the $(d=6)^2$ EFT approximation.
Even for large values of~$M_\mathrm{cut}$, the EFT provides accurate limits, at least for leptoquark masses $\gtrsim 2$~TeV.
Moreover, we see that the $(d=6)^2$ EFT approximation tends to overestimate the constraints on the coupling-over-mass ratio.
While the New-Physics scales probed by the data mostly depend on the considered quark generations, and the energy range where data are collected depends on the flavor of the leptonic final state, we find similar overall patterns for the EFT convergence for all flavor combinations.

\section{Illustration: \texorpdfstring{$b\to c\tau\nu$}{ b -> c tau nu}}
\label{sec:numerical}
%
\begin{figure*}[!t]
\centering
\includegraphics[width=0.32\textwidth]{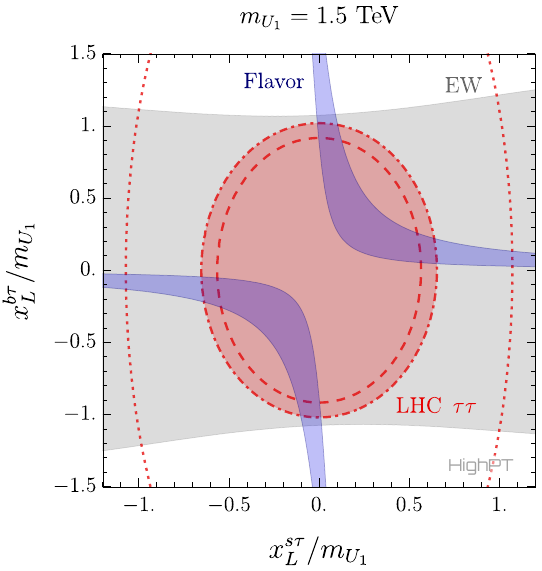}
\includegraphics[width=0.32\textwidth]{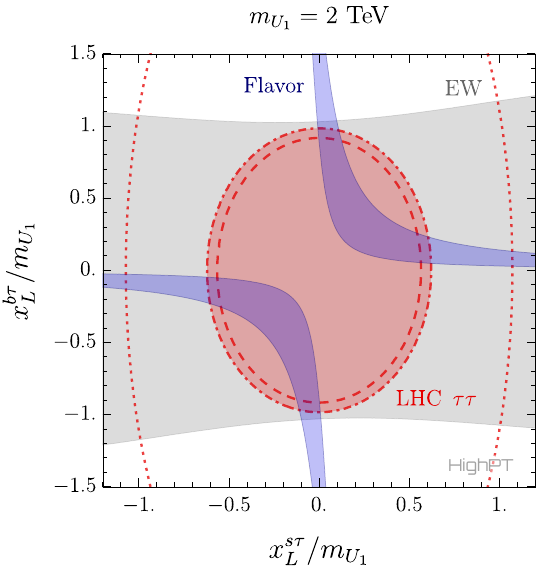}
\includegraphics[width=0.32\textwidth]{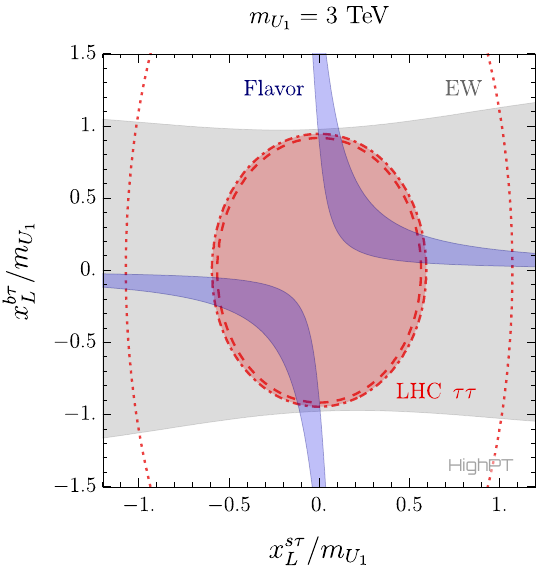}
\caption{Limits on coupling/mass for a $U_1$~leptoquark. The filled red regions are 2$\sigma$~regions coming from Drell-Yan at the LHC using the explicit mediator propagation in {\tt HighPT}~\cite{Allwicher:2022mcg}, while the dotted, dashed, and dot-dashed lines indicate the 2$\sigma$~contours in the SMEFT case with $d=6$ linear, $d=6$ quadratic terms, and $d=8$ quadratic terms, respectively. 
RG~evolution for the flavor and electroweak likelihood is performed taking $m_{U_1}$ as the high scale, i.e., it is different for the three plots.}
\label{fig:U1combinedplots}
\end{figure*}

In this Section, we consider the discrepancies in exclusive $B$-meson decays with an underlying $b\to c\tau\nu$ transition as a concrete example to quantitatively assess the validity of the EFT description of $pp\to\tau\tau$ at the LHC. Drell-Yan processes provide helpful constraints on the New-Physics scenarios proposed to accommodate these discrepancies~\cite{Faroughy:2016osc,Allwicher:2022gkm}, since these low-energy processes occur at tree level in the SM, being only sensitive to scales~$\Lambda$ at most at the $\mathcal{O}(\mathrm{TeV})$~range with current precision~\cite{DiLuzio:2017chi}. 

Our goal is to directly compare the EFT results with the constraints obtained in the full model, for different values of the mediator mass, as in the previous sections, and to study the convergence of the two approaches for large $\Lambda$~values, this time also comparing Drell-Yan bounds to other complementary data sets in the flavor and electroweak sectors.
To do so, we focus again on the $U_1$~vector leptoquark as a benchmark model, which is known to provide a good solution to the $b \to c \ell\nu$ discrepancies~\footnote{See Refs.~\cite{Becirevic:2024pni,Becirevic:2024iyi} for other viable leptoquark scenarios.}~\cite{Buttazzo:2017ixm,Angelescu:2018tyl,Cornella:2019hct,Cornella:2021sby,Angelescu:2021lln}. 

The observables determined at low energies are the ratios $R_{D^{(\ast)}}=\mathcal{B}(B\to D^{(\ast)}\tau \bar{\nu})/\mathcal{B}(B\to D^{(\ast)}l \bar{\nu})$ (with $l=e,\mu$), which have been measured at LHCb and the $B$-factories~\cite{Banerjee:2024znd} 
\begin{align}
    R_D^{\rm exp} = 0.342 \pm 0.026 \,,  \qquad R_{D^\ast}^{\rm exp} = 0.287 \pm 0.012 
\end{align}
with a combined value that is about $3\sigma$ above the average of SM predictions~\cite{Banerjee:2024znd}~\footnote{See also Ref.~\cite{FermilabLattice:2021cdg,Harrison:2023dzh,Aoki:2023qpa} for recent determinations of $B\to D^\ast$ form-factors of the lattice, which have been combined to predict $R_{D^\ast}$, e.g.,~in Ref.~\cite{Martinelli:2023fwm,Bordone:2024weh}.}
\begin{align}
    R_D^{\rm SM} = 0.298 \pm 0.004 \,, \qquad
    R_{D^\ast}^{\rm SM} = 0.254 \pm 0.005 \,.
\end{align}
The leptoquark effect in these observables can be schematically parametrized as
\begin{align}
    \nonumber
    \frac{R_{D^{(*)}}}{R_{D^{(*)}}^{\rm SM}} &\simeq 1 -\frac{v^2}{\Lambda^2} \Re \left( \frac{1}{V_{cb}^*}[\cC_{lq}^{3}]_{3323} + [\cC_{lq}^{3}]_{3333} \right) \\
    &= 1 + \frac{v^2}{2 m_{U_1}^2} \Re \left( \frac{x_L^{23}x_L^{33^*}}{V_{cb}^*} + |x_L^{33}|^2 \right) \,,
\end{align}
where we have again assumed down alignment for the left-handed quark doublet, and kept couplings to the $\tau$~lepton, and the second- and third-generation quarks.
Other relevant constraints on this scenario are posed by the modifications of $Z$~and $W$~couplings to leptons induced by the RG evolution, which are sizable as they are proportional to the top Yukawa coupling~$y_t$ in this case~\cite{Feruglio:2016gvd,Feruglio:2017rjo,Cornella:2018tfd,Allwicher:2022vbf,Allwicher:2023aql}. 
In our case, the biggest effect comes from $Z \nu \bar\nu$ vertex corrections, parametrized by~\cite{Jenkins:2013zja,Jenkins:2013wua,Alonso:2013hga}
\begin{align}
    \hspace{-0.2em}\delta g_L^{\nu} \propto [\cC_{Hl}^{(1-3)}]_{33} (m_Z) &\simeq \frac{6 |y_t|^2}{16\pi^2} \log\frac{m_Z}{\Lambda} [\cC_{lq}^{(1+3)}]_{3333}\,.
\end{align}
For the Electroweak Precision Observables~(EWPOs), we use the inputs from Ref.~\cite{Breso-Pla:2021qoe}.
The running for flavor observables, on the other hand, is negligible for operators with vector structure.
The favored regions from flavor and EWPOs are shown in Fig.~\ref{fig:U1combinedplots} as blue and gray regions, respectively, as a function of the coupling/mass ratio, for three benchmark masses, namely $m_{U_1}=1.5$~TeV (left panel), 2~TeV (center panel) and 3~TeV (right panel).
For each case, the matching scale is taken to be the leptoquark mass~$m_{U_1}$, performing the RG~evolution from this value down to $\mu_{\mathrm{ew}} \approx m_Z$ (hence the slightly different shapes of the electroweak regions for different $m_{U_1}$~values)~\cite{Jenkins:2013zja,Jenkins:2013wua,Alonso:2013hga}.

Turning our attention back to the Drell-Yan constraints shown in red, we observe a good convergence of the EFT (dashed line) toward the leptoquark model (filled red region), with an $\cO(10\%)$~difference between them for $m_{U_1} = 1.5$~TeV.
Notice that the $d=6$ EFT constraints do not change between the three cases, since the only relevant quantity that can be probed is the coupling/mass ratio. 
The EFT description can be further improved by including $d=8$ operators, with a consistent truncation at $\mathcal{O}_{\mathcal{A}}(\Lambda^{-4})$ (dot-dashed red line), which are more important for low $m_{U_1}$ values, to accurately describe the full model.
Already for low leptoquark masses of $m_{U_1}=1.5$~TeV, this truncation order offers an excellent approximation of the full model.
For masses~$\lesssim 1.5$~TeV, we have noticed that the cross section for the last bin of the experimental search~\cite{ATLAS:2020zms} becomes negative when truncating the EFT series at $\cO_{\!\sigma}(\Lambda^{-6})$ on cross-section level, i.e., considering the interference of $d=6$ with $d=8$ operators. 
This highlights the issues associated to cross-section level truncations and further reinforces our statement that it is necessary to truncate the EFT series on amplitude level.
Finally, we note that already at $m_{U_1} = 3$~TeV the differences between the various approaches are completely negligible, in line with the findings of the previous sections.

On the other hand, it appears clear from all three plots that keeping only the linear terms in SMEFT coefficients (dotted line) when computing the cross section never yields accurate results. 
In fact, stopping the expansion at $\mathcal{O}_{\!\sigma}(\Lambda^{-2})$ leads to overly pessimistic bounds. 
This should be compared with the middle and right plots in the bottom of Fig.~\ref{fig:U1clipping}, where the same pattern is found. While one could regard these as conservative bounds, it is worth noticing that this is not always the case, and in most cases the linear dimension-six terms seem to lead to a too restrictive constraint (cf.~again Fig.~\ref{fig:U1clipping}). 

In summary, we find that EFTs can reliably describe Drell-Yan tails for $t$- and $u$-channel mediators with masses above $1.5$~TeV, including the models proposed to address the $R_{D^{(\ast)}}$~anomalies.
We have shown that the EFT description at $\mathcal{O}_{\mathcal{A}}(\Lambda^{-2})$ is rather accurate for these masses and it can be systematically improved by including the $\mathcal{O}_{\mathcal{A}}(\Lambda^{-4})$ contributions.

\section{Summary}
\label{sec:summary}

In this article, we have explored the validity range of the EFT description of Drell-Yan processes and the uncertainties associated with the EFT truncation through a direct comparison between concrete models and their respective EFTs at low energies. We have considered two representative scenarios, namely a heavy $Z^\prime\sim (\mathbf{1},\mathbf{1},0)$ boson and a vector leptoquark $U_1 \sim (\mathbf{3},\mathbf{1},2/3)$, which contribute to these processes via the $s$- and $t$-channels, respectively, and which are matched onto the SMEFT with operators up to dimension $d=8$. 

Firstly, we have computed the analytical expression for the partonic cross section in each of the scenarios, accounting for the propagation of the new mediators. By performing an expansion of the cross section in the ratio $x_V=m_{\ell\ell}^2/m_V^2 < 1$, between the dilepton invariant-mass squared and the squared mass of the mediators, we have shown that the series obtained for the $t$-channel mediator has a faster convergence at low energies than the one obtained for the $s$-channel. This is expected from the resonant nature of the contributions in the former scenario and can be traced back to the coefficients in the power series of the cross section, which appear with the same signs for the $s$-channel propagator, but with alternating signs for the $t$-channel one. These subleading corrections are associated with higher-dimensional operators (i.e.,~beyond $d=6$), which can improve the description of the partonic cross sections as shown in Figs.~\ref{fig:diffcrosssection-Zp} and~\ref{fig:diffcrosssection-U1}.

We reiterate that only considering the interference between the SM and $d=6$ operators [i.e.,~$\mathcal{O}_{\!\sigma}(\Lambda^{-2})$ truncation in our notation] leads to a poor EFT description in most cases, making the inclusion of $(d=6)^2$ terms [$\mathcal{O}_{\!\cA}(\Lambda^{-2})$] necessary.
Furthermore, while the interference term of the $d=8$ operators with the SM amplitude [$\mathcal{O}_{\!\sigma}(\Lambda^{-4})$] only marginally improves the EFT description, we found that $d=8$ squared contributions [$\mathcal{O}_{\!\mathcal{A}}(\Lambda^{-4})$] can have a sizable impact on the cross section if $E/\Lambda$ is not small. In other words, it is preferable to perform the truncation at the amplitude level instead of the cross-section level. This feature is potentially connected to the well-defined properties of high-energy scattering amplitudes, which are governed by analyticity and unitarity (in our case, below the EFT cutoff $E<\Lambda$), which may be lost with an inconsistent truncation on cross-section level. 
Studying $d=8$ effects this way is, of course, only possible if a concrete UV~model is considered. 
Assessing the impact of $d=8$ terms in a model-agnostic manner is less obvious. 
An option would be to fit the $d=6$ Wilson coefficients while marginalizing over the corresponding coefficients of the $d=8$ operators as done in~Ref.~\cite{Allwicher:2022gkm}. 
However, as shown in that reference, the complete decorrelation of the $d=6$ and $d=8$ effects obtained through the marginalization leads to a significant relaxation of the constraints.
Although this can be interpreted as a conservative limit, a considerable improvement can be achieved once a specific model is considered~\cite{Allwicher:2022gkm}.

By comparing the different panels of Figs.~\ref{fig:diffcrosssection-Zp} and \ref{fig:diffcrosssection-U1}, we have also shown that the EFT convergence depends on the couplings convoluted with the initial quark PDFs, providing a better description of the full-model results for small New-Physics couplings and/or small PDFs (i.e.,~heavier quarks). In practice, the range of couplings that are probed by real data depends on the experimental sensitivity, which is usually better for $\ell=e,\mu$~\cite{CMS:2021ctt} than for $\ell=\tau$~\cite{ATLAS:2020zms}, as the experimental searches usually cover higher energies in the former case. These features are illustrated in Fig.~\ref{fig:U1clipping} for the $U_1$~leptoquark, where the EFT description with $d=6$ operators is more accurate and provides better limits for the lighter flavors of quarks and leptons.

Finally, we have also made this explicit comparison for a concrete example motivated by discrepancies between the SM predictions and the experimental determinations of the low-energy $b\to c\tau \nu$ transition. We have considered the vector leptoquark model that was proposed to accommodate this discrepancy, and we have compared the EFT and full-model Drell-Yan bounds on the couplings that are fixed by flavor and electroweak data. We have shown that the EFT bounds based on $d=6$ operators lead to bounds that are $\mathcal{O}(10~\%)$ stronger than the correct ones for a leptoquark of mass $m_{U_1}=1.5~\mathrm{TeV}$, which is currently allowed by direct searches at the LHC~\cite{CMS:2023qdw,ATLAS:2023vxj}.
We have demonstrated that the EFT truncated at $\cO_{\!\cA}(\Lambda^{-2})$, i.e., including $d=6$ operators and truncating on amplitude level, provides a fairly good description of this scenario for masses above $1.5$~TeV, which can be further improved by considering $d=8$ operators with a consistent $\cO_{\!\cA}(\Lambda^{-4})$ truncation. 

While verifying the validity of the EFT approach in general is a delicate problem as it depends on several factors: the specific UV~scenario (resonant or non-resonant), the flavors of all involved particles (particularly in light of PDF-suppression effects and $\tau$-reconstruction issues), the processes considered (kind of process, tree or loop level), etc., the central results of the present work can be succinctly summarized by: 
(\textit{i})~for high-energy observables one should truncate the EFT series on amplitude level, since only this can guaranty the proper analytical structure of amplitudes at high energies, whereas a truncation on cross-section level can lead, for example, to negative cross sections in the high-energy bins; 
(\textit{ii})~while the applicability of EFTs for resonant UV~scenarios is limited to cases of large scale hierarchies~$E \ll \Lambda$, EFTs can offer good approximations for non-resonant scenarios even if the considered energies are not far below the cutoff scale. 
In many cases truncating at~$\cO_{\!\cA}(\Lambda^{-2})$ already provides good approximations in this case, wheres going to~$\cO_{\!\cA}(\Lambda^{-4})$ can offer excellent approximations even for $E\sim\Lambda$.

\appendix

\section{Phase-space functions}
\label{app:lq-formulas}

In this Appendix, we provide the explicit expressions for the phase-space functions defined in Eq.~\eqref{eq:phi-int} through the integral on $\hat{t}\in(-\hat{s},0)$,
\begin{align}
\varphi_1(x) &= \dfrac{3(2+3x)}{2x}-\dfrac{3(1+x)^2}{x^2}\,\log (1+x)\,, 
\\[0.4em]
\varphi_2(x) &= 6+ {3x} - \dfrac{6(1+x)}{x}  \log(1+x)\,.
\end{align}
These expressions can be expanded for $0<x<1$, leading to the expressions in Eq.~\eqref{eq:phi-series}.

\vspace*{0.4cm}

\section*{Acknowledgments}

We warmly thank Florentin Jaffredo for collaboration at the early stages of this project. This project has received support from the European Union’s Horizon 2020 research and innovation programme under the Marie Skłodowska-Curie grant agreement N$^\circ$~860881-HIDDeN and N$^\circ$~101086085-ASYMMETRY and the IN2P3 (CNRS) Master Project HighPTflavor. The work of DAF~is supported by the US Department of Energy under grant DOE-DE-SC0010008.
FW acknowledges the support of the Deutsche Forschungsgemeinschaft (DFG, German Research Foundation) under grant 396021762 – TRR~257: \textit{Particle Physics Phenomenology after the Higgs Discovery}.
LA~also acknowledges funding from the Deutsche Forschungsgemeinschaft under Germany’s Excellence Strategy EXC 2121 “Quantum Universe” – 390833306, as well as from the grant 491245950. MM~acknowledges the support of Fundação de Amparo à Pesquisa do Estado de São Paulo (FAPESP) under the grant numbers 2022/11293-8 and 2024/04246-9.


\end{document}